# Greenhouse gas emissions in the European Union-28. A multilevel club convergence study of the Emission Trading System and Effort Sharing Decision mechanisms

María José Presno; Manuel Landajo; Paula Fernández González

University of Oviedo

**Abstract**

The European Union is engaged in the fight against climate change. A crucial issue to enforce common environmental guidelines is environmental convergence. States converging in environmental variables are expected to be able to jointly develop and implement environmental policies. Convergence in environmental indicators may also help determine the efficiency and speed of those policies. This paper employs a multilevel club convergence approach to analyze convergence in the evolution of greenhouse gas (GHG) emissions among the European Union (EU)-28 members, on a search for countries transitioning from disequilibrium to specific steady-state positions. Overall convergence is rejected, with club composition depending on the specific period (1990-2017, 2005-2017) and emissions categories (global, Emission Trading System (ETS), Effort Sharing Decision (ESD)) analyzed. Some countries (e.g. the United Kingdom and Denmark) are consistently located in clubs outperforming the EU's average in terms of emissions reductions, for both the whole and the most recent periods, and for both ETS and ESD emissions. At the other end, Germany (with a large industrial and export basis), Ireland (with the strongest gross domestic product (GDP) growth in the EU in recent years) and most Eastern EU members have underperformed after 2005, almost reversing their previous positions when the study begins in year 1990. Member States show large mitigation potentials in both ETS emissions (e.g., in the aviation sector) and the multiplicity of sectors (especially road transport) covered by the ESD. The presence of multiple convergence clubs, reinforced by persistence of the results across the various periods and sectors analyzed, may also indicate the convenience of more differentiated, specific policies and regulations with a view to reducing heterogeneity among member states. Innovation in production and business systems, the promotion of green energy, collaboration and efficient technology transfer between countries, a transparent regulatory framework, and the promotion of green attitudes are key measures for reducing GHG emissions, especially in those countries with the worst performance.

**NOMENCLATURE**

$y_{it}$ logarithm of the GHG emissions index (with basis 1990 or 2005) for country $i = 1, \ldots, N$ and year $t = 1, \ldots, T$

$N$ number of countries

$T$ sample size

$\delta_{it}$ transition parameter

$\mu_t$ common growth component

$h_{it}$ relative transition parameter

$H_t$ cross-sectional variance of $h_{it}$

$[rT]$ integer part of $rT$

$\hat{b}$ least squares estimator for parameter $b$

$\alpha$ speed of convergence

$t_{\hat{b}}$ one-sided $t$ statistic of the null hypothesis $b \geq 0$

# 1. Introduction

The United Nations 2030 Agenda for Sustainable Development includes a list of 17 Sustainable Development Goals (SDG) and 169 targets seeking to build on the Millennium Development Goals. Goal 13 particularly refers to climate action, emphasizing the need of taking urgent measures to fight climate change and its impact.

In the specific case of the European Union (EU), some policies included in the EU Sustainable Development Strategy (European Council, 2001) have also been focused on climate change, with progress being assessed through so-called Sustainable Development Indicators (SDI). Another relevant background is the Europe 2020 Strategy (European Council, 2010), which puts forward priorities to turn Europe into a smarter, more sustainable and inclusive place to live. That strategy poses 3 specific targets for climate change and energy policy (namely, as compared with their 1990 levels, (i) reducing greenhouse gas (GHG) emissions by 20%, (ii) increasing to 20% the share of renewables in final energy consumption, and (iii) moving towards a 20% increase in energy efficiency) to be reached by year 2020. Additionally, in May 2017, in response to the United Nations 2030 Agenda, the European Statistical System Committee adopted the SDG indicator set, which comprises 100 indicators structured in 17 SDGs. Most of those indicators stem from pre-existing indicator sets (such as the EU SDI, Europe 2020 headline indicators, and the set of impact indicators for the Strategic Plan 2016–2020) already employed for monitoring long-term EU policies.

The European Commission monitors the "climate action" goal by analyzing three sub-themes, namely 'climate mitigation', 'climate impacts', and 'climate initiatives'. The first one focuses on fighting climate change through reductions in GHG emissions and promotion of less carbon-intensive energy. Climate impacts refer to the effects of climate change, and climate initiatives consider specific actions aiming at fighting climate change.

In this paper we shall focus on GHG emissions. This is usually the main indicator employed in order to track the success of mitigation measures against climate change. Certainly, an analysis of the other two sub-themes would also be of interest, but several issues (including the lack of statistical data disaggregated for all EU Member States and the short length of the time series available for the analysis) led us to concentrate on GHG emissions for the EU-28 group of countries.

The European Union has some specific features that make it an interesting case study. First, the European Commission has set up a goal of becoming the first climate neutral continent, with net zero emissions. The EU constitutes a diverse group of countries (in terms of both their economic and cultural antecedents and their climatic and social characteristics) sharing the common overall objectives of attaining a reduction of 20% in GHG emissions by 2020 with respect to 1990 levels. This would be the first step towards a more ambitious goal of at least a 55% drop by 2030 from 1990 levels, and well on the way to reaching climate neutrality by the horizon of year 2050. The EU-28 has committed itself, on both the financial and regulatory fronts, to reach those goals. In this regard, the €1 trillion plan outlined in the European Green Deal has among its aims reviewing current legislations on their climate merits and introducing new laws and regulations on issues like circular economy, renovation of buildings, biodiversity, farming and innovation. The EU is also among the most proactive participants in international organizations involved in fighting climate change.

Our main goal in this paper is the analysis --for several aggregation levels and time spans-- of GHG emissions convergence in the EU-28, with a view to drawing useful environmental guidelines. Several possibilities exist, ranging from strict (so-called *overall*) convergence to divergence, including the interesting intermediate case when several groups of countries (so-called *convergence clubs*) approach different equilibria. We aim at finding groups of EU members that have evolved from disequilibrium to specific steady-state positions, with those transitions being the result of factors like the potential presence of affine characteristics in their growth processes, changes in the composition of their energy production mix (i.e., from non-renewables to renewables), and specific legislations and energy consumption patterns.

The interest of analyzing convergence in GHG emissions derives from such matters as the Environmental Kuznet´s Curve (EKC), global mitigation efforts by organizations and governments to stop climate change, and the relevance of convergence on explaining both pollution emissions and concentrations (Stern, 2017). As for the EKC (e.g., Erdogan and Okumus, 2021), a classical paper by Strazicich and List (2003) may suggest a potential connection, with convergence in environmental degradation being implied by the EKC concept.

Environmental convergence is also crucial because of its policy implications (Aldy, 2006; Burnett, 2016; Apergis *et al.,* 2017), as it is expected that those states converging in terms of their environmental variables will also be able to more effectively implement common environmental policies. At the same time, convergence in environmental indicators might help determine the efficiency and speed of environmental policies (Bilgili and Ulucak, 2018).

Many previous studies in the field of convergence have focused on stochastic convergence, with unit root/stationary testing methods being employed. Instead, we shall use the club convergence approach of Phillips and Sul (2007, 2009) which, in addition to testing for convergence, enables a more flexible and systematic analysis of club convergence issues, including routines for club identification, estimation and testing. More precisely, their procedure makes it possible to identify both groups of countries converging to different equilibria and specific countries that diverge from the rest of the group. This opens the door to the possibility of analyzing potential relationships between clubs of convergence and certain economic characteristics, as well as identifying potential reasons for divergence.

Unlike prior studies (e.g., Apergis and Payne, 2017; Ulucak and Apergis, 2018; Emir *et al.*, 2019) where clubs of convergence for total and per capita environmental variables were analyzed, we conduct our study for the index itself. This choice was motivated by the fact that the objectives of most policies in the field are formulated in relative (i.e., as percent increases/decreases in the magnitudes) instead of absolute terms.

Our analysis proceeds in two stages: first an overall study for the whole 1990-2017 period will be carried out, and then a separate analysis is conducted for the 2005-2017 subperiod. Our intention when singling out the latter period was to more closely monitor the process of convergence in recent years and, more specifically, to analyze the performance of each Member State under two instruments of the EU's 2020 climate and energy package, namely the Emission Trading System (ETS) and the Effort Sharing Decision (ESD). Certainly, a convergence process may take a long time, so limitations in data availability in the interim may pose hardly avoidable limits on the scope of empirical studies and reliability of conclusions. In this regard, our analysis would share the potential drawbacks of any approach that relies on real time monitoring of a long run process.

The remainder of the paper is organized as follows: Section 2 includes a review of the literature in the field. Section 3 briefly describes the methodology of club convergence analysis. Section 4 includes our main results with a discussion. Finally, Section 5 outlines some conclusions.

## 2. Literature Review

From a historical standpoint, the convergence concepts employed nowadays to analyze the evolution of environmental magnitudes were originally developed in studies on income growth, where several convergence concepts have been proposed, including beta convergence (introduced by Baumol, 1986), sigma convergence (which traces back to Barro and Sala-i-Martin, 1990), stochastic convergence (developed by Quah, 1990 and Carlino and Mills, 1993), and club convergence.

The neoclassical growth model proposed by Solow (1956) predicts that in the long run --under technological homogeneity and identical preferences-- cross-country differences in per capita real income tend to decrease as each country approaches its balanced growth path, with overall convergence eventually occurring. Azariadis and Drazen (1990) and Galor (1996) showed that multiple equilibria are also possible, with those countries having similar structural characteristics (e.g., government policies and production technology) but different initial conditions converging to different steady-state equilibria. Therefore, for a group of similar economies, a common growth path could only be attained provided that their initial conditions are all in the basin of attraction of the same steady-state equilibrium. Those countries would conform a convergence club.

Conceptually, the idea of stochastic convergence in carbon emissions implies that shocks in the logarithm of per capita carbon emissions, relative to the average of the group, are trend-stationary (or mean-stationary in the deterministic convergence case). Hence, the potential presence of a unit root would indicate that the effect of shocks is permanent in nature and the series diverges from the group average. Unit-root and stationarity (both individual and panel) tests are applied to detect that pattern.

Following Carlino and Mills (1993), stochastic convergence occurs when both stationarity around a trend and beta-convergence simultaneously hold.

The seminal contribution on convergence in $CO_2$ emissions is due to Strazicich and List (2003), who find significant evidence of stochastic convergence in 21 OECD countries. Thereafter, several papers have addressed the same problem, for different periods and groups of countries, and employing a highly diverse set of statistical tools that includes unit root and stationarity tests allowing for structural breaks (e.g., Lee and Chang, 2009), panel tests (e.g., Westerlund and Basher, 2008; Apergis *et al*., 2017), nonlinear models (Camarero *et al*., 2011; Yavuz and Yilanci, 2013; Presno *et al*., 2018), and panel stationarity tests with trigonometric functions to estimate smooth changes in the trend of the series (Erdogan and Acaravci, 2019), among many others (see Erdogan and Okumus, 2021, for an up-to-date review of the relevant literature).

As for club convergence, one of the first applications to environmental variables is due to Panopoulou and Pantelidis (2009), who analyzed a database of 128 countries, concluding that carbon emission convergence occurs along with income convergence. Herrerias (2013), in a study conducted on a broad panel including both developed and developing countries, with carbon dioxide emissions separated by energy source (namely coal, petroleum, and natural gas), finds evidence of convergence for a large portion of the countries. Camarero *et al*. (2013) analyze OECD countries and $CO_2$ emission intensity, whereas Wang *et al*. (2014) focus on Chinese provinces, and Burnett (2016) and Apergis and Payne (2017) study U.S. states. At the European Union level, Emir *et al.* (2019) focus on carbon intensity in the EU-28 countries, also analyzing convergence among the EU-15 and new EU members that joined in after 2004 (for both periods, before and after the incorporation of the new members), finding no evidence of convergence. Morales-Lage *et al*. (2019) study per capita $CO_2$ emissions in the 1971-2012 period, with a focus on the energy subsectors (namely power generation and heating, manufactures and construction, transportation, and other minor fuel combustion), observing that core European countries (France, the Netherlands, Germany and the United Kingdom) are included in the top performing clubs for all the subsectors.

Although most studies analyzing environmental degradation have focused on $CO_2$ emissions, Rees and Wackernagel (1996) and Wackernagel and Rees (1998) propose as an alternative the concept of ecological footprint, which provides a multi-dimensional framework that includes six sub-components (namely cropland, grazing land, fishing grounds, forest land, built-up land, and carbon footprints) and takes the environmental pressures of human activities into account.

Some recent papers have analyzed convergence of the ecological footprint for several country panels and sample sizes, and employing different methodologies. For instance, Ulucak and Apergis (2018), Haider and Akram (2019) and Solarin *et al.* (2019) apply clubs of convergence; panel stationarity or unit root tests are employed, among others, by Ulucak and Lin (2017), Yilanci *et al.* (2019), and Okumus and Erdogan (2019); some studies (e.g., Bilgili and Ulucak, 2018; Erdogan and Okumus, 2021) make a combined use of both methodologies.

As commented above, this paper addresses a convergence analysis on the evolution of GHG emissions in the EU-28. We adopt a multi-level standpoint, with two periods (and base years; respectively, 1990 and 2005) and several categories (global, ETS, ESD emissions) considered simultaneously. To the best of our knowledge, this would be the first study applying the club-of-convergence approach to the GHG index, with a disaggregation by both ETS and ESD.

## 3. Methodology

We consider a panel data set obeying the following single factor model:

$$y_{it} = \delta_{it}\mu_t, \text{ for all } i, t \tag{1}$$

where $y_{it}$ denotes the logarithm of the GHG emissions index (with basis 1990 or 2005) for country $i = 1, \ldots, N$ and year $t = 1, \ldots, T$, $\delta_{it}$ (the transition parameter) is a time-varying idiosyncratic element, and $\mu_t$ (a common growth component) captures some deterministic/stochastic trending behavior in $y_{it}$. Thus, $\delta_{it}$ may be interpreted as the individual "economic distance" between $\mu_t$ and $y_{it}$.

In order to model the transition elements $\delta_{it}$, Phillips and Sul (2007) construct the "relative transition parameter" as follows

$$h_{it} = \frac{y_{it}}{N^{-1}\sum_{i=1}^{N} y_{it}} = \frac{\delta_{it}}{N^{-1}\sum_{i=1}^{N} \delta_{it}} \tag{2}$$

$h_{it}$ traces out the individual transition path over time for economy *i* in relation to the panel average.

Phillips and Sul (2007) stress some properties of $h_{it}$, including: (1) by definition, the cross-sectional mean of $h_{it}$ equals one, and (2) under convergence, the relative transition parameters $h_{it}$ converges in probability to 1 for all $i$ as $t \to \infty$, and the cross-sectional variance of $h_{it}$, namely

$$H_t = N^{-1} \sum_{i=1}^{N} (h_{it} - 1)^2 \qquad (3)$$

would converge to zero.

Based on these considerations, Philips and Sul (2007) propose the so-called *logt* convergence test, which involves estimating the following OLS regression:

$$log\left(\frac{H_1}{H_t}\right) - 2 \log(\log(t)) = a + b \log(t) + \varepsilon_t \qquad (4)$$

where $t=[rT], [rT]+1, …, T$, for $r=0.3$ in the case of small/moderate sample sizes ($T \leq 50$), with $[rT]$ being the integer part of $rT$.

Under the null of convergence, the least squares estimator for parameter $b$ converges in probability to $2\alpha$ (the scaled speed of convergence), so the null hypothesis can be tested through a one-sided $t$ test of the null $b \geq 0$, which is rejected at 5% significance if $t_{\hat{b}} < -1.65$.

Since $b = 2\alpha$, it is readily checked that the case $b \geq 2$ (i.e., $\alpha \geq 1$) implies convergence in levels (so-called absolute convergence), whereas $0 \leq b < 2$ means that the growth rates converge over time (conditional convergence).

Rejection of the null hypothesis of convergence for the whole panel certainly does not rule out the possibility of convergence for specific clusters or subgroups. To investigate that prospect Phillips and Sul (2007) propose a four-step clustering algorithm that allows researchers to endogenously identify clubs of convergence in the panel.

As their initial algorithm tends to over-estimate the number of clubs, Phillips and Sul (2009) propose to merge the starting groups into larger clubs by using the *logt* convergence test. The same test may also be used to investigate the possibility of *transitioning* between clubs (i.e., clubs slowly converging to one another, or some members of a club moving towards a contiguous club).

## 4. Results and discussion

In this Section we analyze convergence in GHG emissions from two different time origins (respectively, years 1990 and 2005). As commented above, the latter will allow us both to study the evolution (using as a reference a more recent period when most states in the panel already belonged to the EU-28) and to separate emissions under the ETS and ESD mechanisms. (Table A.1 in the Appendix provides some historical background on the EU's member states and their emissions.)

### 4.1. Convergence clubs for the 1990-2017 period

The European Commission evaluates GHG emissions through an index having 1990 as the base year, with its data source being the European Environment Agency (EEA). The index entails total man-made emissions (including international aviation) of the so called 'Kyoto basket'. According to that indicator, GHG emissions in the EU shrank by 21.7% between 1990 and 2017, clearly decoupling from the evolution of GDP (that increased by 57%) and putting the EU on track to surpass its 2020 target. Overall, the main causes of that reduction would include an increasing use of less intensive fuels, improvements in energy efficiency and a growing use of renewables, in addition to other factors as structural changes in the economy and the economic recession following the financial crisis of 2008. Nevertheless, several studies suggest that the pace of emission reductions would tend to slow down after year 2020. This would be potentially problematic, as continuing GHG reductions at a slower pace would prove insufficient to achieve the EU´s target by year 2030, particularly after the slight rebound (+0.6%) in 2017 with respect to year 2016.

Beyond the observed reductions in overall GHG emissions, it is important to assess whether overall convergence occurs in the EU or, on the contrary, some countries tend to form convergence clubs regardless of common EU environmental policies and objectives. At a glance, a certain degree of heterogeneity is detected in the evolution of emissions, with some countries (e.g., Cyprus –with a 55.7% increase- and, to a lesser extent, Portugal, Spain, Ireland, Malta, and Austria) increasing their emissions in 2017 in comparison with 1990, whereas other states (e.g., Romania and the Baltic Republics of Lithuania and Latvia) underwent cuts of more than 50%.

Since the index has 1990 as its base year, our analysis will be carried out for the 1991-2017 period. As a matter of fact, the clustering algorithm removes the first 30% of observations (which helps mitigate the sensitivity of the test to initial effects), so all calculations for the *logt* test statistic actually refer to the 1999-2017 subperiod, when the dispersion associated with the initial effect would hopefully have dissipated.

The results of the *logt* convergence test appear in Table 1 below. Figure 1 displays the average relative transition curves for each club (reflecting the change in GHG emissions of each club in relation to the average level).

**Table 1. GHG emissions, 1990-2017.**

| | | Initial classification | | | Test of club merger | | | Transition | | |
|---|---|---|---|---|---|---|---|---|---|---|
| | | $t_{\hat{b}}$ | $\hat{b}$ | $\hat{a}$ | | $t_{\hat{b}}$ | $\hat{b}$ | | $t_{\hat{b}}$ | $\hat{b}$ |
| | | -13.953* | -0.504 | | | | | | | |
| Club 1 | CY, MT | -0.579 | -1.045 | -0.522 | | | | | | |
| Club 2 | PT, ES, IE, AT, SI, LU, PL, EE | 0.675 | 0.092 | 0.046 | Club 1+2 | -2.193* | -0.166 | MT plus PT, ES, IE, AT | -0.745 | -0.348 |
| Club 3 | EL, NL, FR, IT, FI, HR, BG | 0.779 | 0.132 | 0.066 | Club 2+3 | -0.210 | -0.021 | SI, LU, PL, EE plus EL, NL, FR, IT | 1.872 | 0.273 |
| Club 4 | BE, SE, DE, DK, HU, CZ, UK, SK, RO, LV, LT | 0.320 | 0.017 | 0.008 | Club 3+4 | -3.004* | -0.203 | FI, HR, BG plus BE, SE, DE, DK, HU, CZ | 1.399 | 0.342 |

* Rejection of the null of convergence at 5% significance.

Belgium (BE), Bulgaria (BG), Czech Republic (CZ), Denmark (DK), Germany (DE), Estonia (EE), Ireland (IE), Greece (EL), Spain (ES), France (FR), Croatia (HR), Italy (IT), Cyprus (CY), Latvia (LV), Lithuania (LT), Luxembourg (LU), Hungary (HU), Malta (MT), Netherlands (NL), Austria (AT), Poland (PL), Portugal (PT), Romania (RO), Slovenia (SI), Slovakia (SK), Finland (FI), Sweden (SE), United Kingdom (UK).

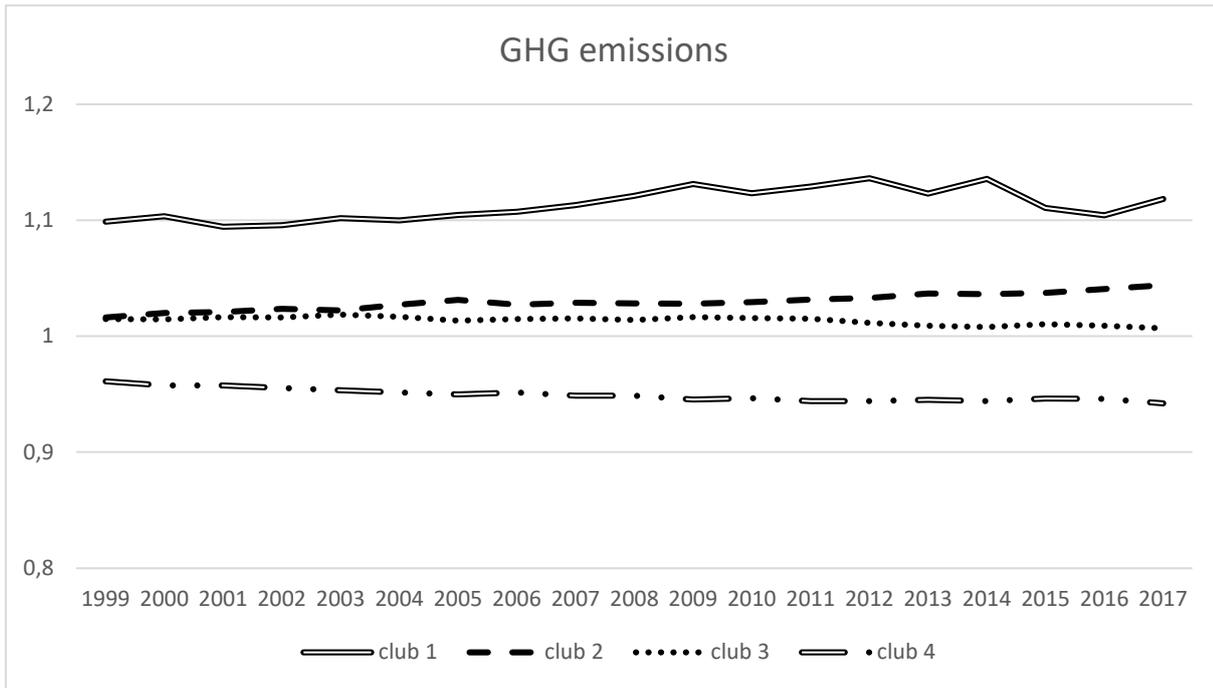

**Figure 1. Relative transition paths across GHG emissions clubs. 1990-2017**

The *logt* test indicates that the null of overall convergence is rejected at 5% significance, so the EU28 countries would not have converged during the study period to a single equilibrium state in their GHG emissions. In order to check for the potential presence of clubs of convergence we also implemented the algorithm by Phillips and Sul (2007), which identifies the following four clubs:

- Club 1 is the highest emissions club. It is formed by two islands, Cyprus and Malta, whose electricity generation has been dominated by emission intensive fossil fuels. It is noteworthy that Malta reduced total GHG emissions by 29% between 2014 and 2016, primarily due to a sharp decrease in its energy supply sector; however, emissions bounced back again in 2017. Results by Emir *et al*. (2019), in their analysis of carbon intensity for the EU-28, also stress the poor performance of these two states.

As seen in Figure 1, Club 1 has a relative transition path with values above 1 and exhibits an upward trend, somewhat alleviated by the effect of the recent evolution of emissions in Malta. It is also remarkable that the point estimate for *b* is negative, though not significantly different from zero; so the null of convergence is not rejected at 5% significance, which suggests that (maybe as a consequence of Malta´s recent performance) those countries would conform to a weak convergence club.

- Club 2 is integrated by a group of EU countries (namely, Spain, Portugal, Ireland, and Austria) that, along with Cyprus and Malta, increased their total emissions between 1990 and 2017. Other nations also included in the club are Luxembourg and three Eastern states (Slovenia, Poland, and Estonia). The case of Estonia may appear paradoxical at first glance, since that country decreased its emissions by 48% in the period, but a recent rebound may have caused its inclusion in the club. The relative transition path in Club 2 (Figure 1) reveals a slightly growing trend with transition values exceeding 1 (i.e., above the mean of the EU-28 group). Regarding speed of convergence, the club exhibits so-called conditional convergence ($\hat{\alpha} = 0.046$), meaning that growth rates converge over time.

- Club 3 includes a group of long-time EU Member States (France, Italy, the Netherlands, Finland, and Greece), plus two more recent Central/Eastern members (Croatia and Bulgaria). In the case of France, its energy policy has been based largely on use of nuclear energy for power generation, whereas green growth is a priority in Finland. As for Italy and Greece, emissions decreased significantly since 2007, probably as a consequence of the economic crisis. Finally, the relative transition path in Club 3 reveals a flat pattern with values around 1 (i.e., slightly above the mean of the EU-28), as well as conditional convergence.

- Club 4 is the largest club, including Germany, the United Kingdom, Sweden, Denmark, and Belgium. Germany, despite being the highest emitter in the EU (with 21% of total emissions) has experienced significant reductions due to such factors as increases in efficiency of its power and heating plants, economic restructuring after German reunification, a reduction in the carbon intensity of fossil fuels (with the switch from coal to gas), and a sharp increase in renewable energy use and waste management measures. Meanwhile, the United Kingdom -the second largest emitter in the EU- reduced emissions, mainly as a result of liberalization of its energy markets and the fuel switch from oil and coal to gas in its electricity production, in addition to the use of more efficient combined cycle gas turbine stations and decreases in its iron and steel productions. As for Denmark and Sweden, it is well known that those countries emphasize the use of clean technology and green growth. In Belgium nuclear energy has a significant weight in its energy mix.

Club 4 is also integrated by some Central/Eastern European countries as the Czech Republic, Hungary, and Slovakia (all of them increasing the share of nuclear energy in their energy mix), plus three of the states with the largest reductions observed in the study period, namely Romania, Latvia and Lithuania. In the case of Lithuania, it is remarkable that --in spite of reducing to zero the weight of nuclear resources in its gross inland consumption and electricity generation-- the country experienced the largest reduction in GHG emissions in the period among all EU-28 nations.

It is also observed that most Central/Eastern countries are included in the top performance clubs. When interpreting this result, it must be taken into account that all those nations had high emission levels at the beginning of the period and underwent substantial restructuring of their economies in the late 1980s and early 1990s. Besides that, those states exhibit large differences both in their efforts and in the evolution of their emissions throughout the period, as evidenced by the relatively heterogeneous performance of the various convergence clubs to which they are assigned.

It is also remarkable that, with the exceptions of Poland and Estonia, all the members of the Council of the Baltic Sea States (CBSS) are allocated to Clubs 3 and 4. CBSS countries keep among their long-term priorities a green and low-carbon economy, the promotion of green and sustainable technologies, actions to adapt to climate change and improve the capacity of resilience of ecosystems, and initiatives oriented to protect the ecosystem and biodiversity of the Baltic Sea region.

Phillips and Sul (2009) propose using their *logt* test to investigate the possibility of merging clubs, with a view to avoiding over-estimation of the number of groups. Results in Table 1 also explore potential unions of clubs, concluding that it is possible to combine Clubs 2 and 3, although the resulting value of the test statistic would turn negative, so weaker evidence of convergence would be obtained. We also searched for signs of so-called transitioning between clubs (i.e., clubs that slowly converge to one another, or countries in a club with a tendency to be in transition towards a contiguous group). In this regard, strong evidence of transition -between Clubs 2 and 3, and Clubs 3 and 4- is observed.

The specific composition of each of the above convergence clubs may be put into relation to variables such as growth, energy consumption, and the energy mix. We shall address these connections in Subsections 4.1.1 and 4.1.2 below.

**4.1.1. Growth, energy consumption and emissions**

The EKC hypothesis postulates the existence of an inverted U-shaped relationship between economic growth and environmental degradation (Grossman and Krueger, 1991, 1995). Additionally, the hypothesis of the N-shaped EKC (e.g., Shafik and Bandyopadhyay, 1992; Selden and Song, 1994) postulates a third stage marked by high income levels and low growth rates, with the economy beginning to suffer increases in technical obsolescence that eventually result in a positive relationship between income level and pollution.

With a view to comparing GHG and GDP, Table A.2 in Appendix reports the evolution of the latter. Among the states that experienced stronger (e.g., more than 60%) growth along the period, we find all the members of Clubs 1 and 2 (i.e., those countries having emissions above the mean of the UE-28), with the exception of Portugal. However, some decoupling between growth and emissions is also observed. Indeed, among the countries having stronger growth there are some members of Clubs 3 and 4, like Sweden, the Netherlands, and the United Kingdom (with GDP and population growing simultaneously in all of them throughout the study period), together with the former communist states of Slovakia, the Czech Republic, Romania, and Bulgaria.

Studies on EKC have also researched the relationship between energy consumption and environmental pollution (e.g., Grossman and Krueger, 1991; Acaravci and Ozturk, 2010; Balsalobre-Lorente and Shahbaz, 2016). In this regard, with the only exception of Finland, all the eight countries that experienced stronger growth in energy consumption are allocated to Clubs 1 and 2. Moreover, it is observed that strong GDP growth does not necessarily come along with large increases in energy consumption. Among the older member states, we observe this decoupling in Germany and especially in the United Kingdom. In these two countries GDP (and population) grew up strongly whereas consumption and emissions dropped. This brings about an additional factor: efficiency. In this vein, it is remarkable that the *Combined energy efficiency scoreboard* (by ODYSSEE) in 2016 places both the United Kingdom and Germany among the top five countries in terms of energy efficiency (Lopez *et al*., 2018). Additionally, Fernandez Gonzalez *et* al. (2014, 2015) remark the strong effects of fuel mix (in the case of the United Kingdom) and energy efficiency (in the largest economies, particularly Germany) on reducing emissions.

### 4.1.2. The energy mix in gross inland consumption and electricity generation

Efficiency and change towards less energy intensive sectors are relevant in order to reduce energy consumption and emissions. However, some countries --as Belgium, Denmark, Sweden, and Finland-- increased consumption while reducing their emissions, which points out that other key factors in reducing emissions are employment of less intensive fuels and the growing use of renewables (also see Camarero *et al*., 2013). More precisely, the Energy Union strategy highlights the use of renewable energies as a part of the efforts required for the decarbonization of the energy system, this being a key element in energy policy as it allows both reducing the dependence on fuel imported from non-EU countries and decoupling energy costs from oil prices. In order to achieve the objective of increasing the share of renewable energies in final energy consumption to 20% by year 2020, the Renewable Energy Directive sets binding national targets (ranging from 10% -Malta- to 49% -Sweden-). Additionally, according to the 2030 Climate and Energy Policy Framework, the share of renewables shall further increase to at least 32%.

Since 2004, the RES indicator quantifies the share of renewable energy consumption in gross final energy consumption. That indicator shows that in the 2004-2017 period the use of renewable energy continuously increased in the EU, with its share almost doubling since year 2004 (when renewables covered only 8.5% of gross final energy consumption). Certainly, two main drivers of that increase were the support schemes and obligations linked to renewable energies and falling renewable energy system costs.

Although all the EU countries increased their share of renewable energy in final energy consumption from 2004 on, there has been large variability among countries, depending on both the renewable sources available and the financial and regulatory support provided. Thus, Sweden was the leader in 2017 with a share of 54.5%, followed by Finland and Latvia with respective shares around 40%. At the other end we find the Netherlands and Luxembourg, with shares around 6% (Table A.3).

According to RES data, eleven countries (the Baltic Sea States of Sweden, Finland, Denmark, Estonia, Lithuania, and Latvia, plus Slovenia, Croatia, Romania, Austria, and Portugal) exceeded the mean of the EU RES values in the 2004-2017 period. It is remarkable that many of them are in Clubs 3 and 4. Other states included in Clubs 3 and 4, but not at the forefront of the RES indicator, are France, Bulgaria, the Czech Republic, Slovakia, and Hungary. All of them lifted their percentages of nuclear in gross inland consumption and also exhibit high percentages of that source in their gross electricity generation. Romania has combined increases in nuclear and renewable sources. In the case of Sweden, in spite of maintaining a considerable weight of nuclear energy in its mix, that share has been reduced in favor of renewable sources.

Germany and the United Kingdom are also in Clubs 3 and 4. Although those countries certainly do not lead the RES indicators, they did carry out a restructuring of their energy systems, transferring parts of their gross inland consumption and electricity generation from solid fossil fuels to natural gas, renewables, and biofuels. In particular, the United Kingdom multiplied its RES indicator by more than nine between 2004 and 2017. A similar behavior is observed in Belgium, that retains a high percentage of nuclear energy. Finally, Italy and Greece replaced oil and petroleum products and solid fuels with less intensive natural gas and renewables sources.

Turning now to Club 2, it is observed that Austria, Portugal and Slovenia showed high RES values in 2017. However, those countries already had considerable RES levels in 2004, so a slow development of renewables follows for them in recent years. Ireland, Luxembourg, Poland, and Slovenia are also among the countries that need to step up efforts to comply with the average indicative trajectory from the Renewable Energy Directive for the 2017-2018 period. Their slow growth in renewables adds to the fact that most countries in Club 2 do not consume or generate electricity from nuclear sources or have even reduced it in recent years (e.g., Spain and Slovenia). However, a majority of those countries have increased their shares in natural gas (which is less emissions intensive, but still high; see Table A.4 in Appendix) while keeping high oil and petroleum products shares.

It is also remarkable that most countries in Club 2 have suffered large increases in their emissions from the transport sector. Transport almost amounts to a quarter of EU GHG emissions, being the second largest emitter in the EU after energy industries. Furthermore, that sector has not experienced the same gradual decline in emissions as others: in 2017, GHG emissions from transport were 19.2% higher than in 1990, with the largest increases registered in Poland (203%), Ireland (133%), Luxembourg (116%), Slovenia (103%), Malta (92%), Austria (74%), Croatia (71%), Cyprus (69%), Portugal (68%), the Czech Republic (63%), and Spain (51%).

As a rule, the EU countries have maintained a high share of oil and petroleum products in their consumption. The use of oil products has decreased in the industry and residential sectors in recent years, but still makes up 94% of final energy consumption in the transport sector. More precisely, around 82% of EU's final energy consumption in the transport sector can be attributed to road transport, with aviation having a growing share of overall transport energy consumption in recent years. Thus, strengthening the use of renewable energy sources in the transport sector seems vital in order to control GHG emissions and reduce dependence from third countries.

**4.2. The 2005-2017 period. Analysis of ETS and ESD emissions**

As commented above, it is important to conduct the above analysis separately for the subperiod from 2005 on, since some of the countries in the study had recently joined the EU, and some of the policies, mechanisms and objectives on climate change have been formulated in the years thereafter. This perspective will also allow us to separately analyze emissions under two instruments of EU's climate and energy policy, namely ETS and ESD.

Table 2 and Figure 2 (panel A) below report the results. When only the evolution since 2005 is considered, the null of convergence keeps being rejected, but the composition of certain clubs changes. With the only exception of Romania, none of the newer Member States is now among the top performing countries. Indeed, the worst evolution is observed for some former Communist states, along with Ireland. Briefly, and for ease of comparison, Table 3 includes the classification of each country for each reference year (respectively, 1990 and 2005). The most abrupt changes in club membership occur in Latvia, Germany, and to a lesser extent Lithuania, which move from groups with emissions below the mean of the EU-28 to clubs exhibiting transition paths slightly above 1 and growing trends. Additionally, the inclusion of Ireland in Club 1 seems to be motivated by its evolution since 2015, with sharp GDP growth accompanied by mild decreases in energy consumption. At the other end, Italy and Greece shift from clubs with transition paths slightly over 1 to having below-average evolution. Those countries suffered major setbacks in both energy consumption and GDP (Table A.2) as a consequence of the economic crisis in that period. Denmark, the United Kingdom and Romania remain in the best-performing clubs, regardless of the reference year employed. Romania, despite its high growth, exhibits one of the sharpest reductions in gross inland energy consumption. In the case of the United Kingdom that decrease is even more impressive.

**Table 2. GHG emissions, 2005-2017.**

| | | Initial classification | | | Test of club merger | | | Test of club merger | | |
|---|---|---|---|---|---|---|---|---|---|---|
| | | $t_{\hat{b}}$ | $\hat{b}$ | $\hat{\alpha}$ | | $t_{\hat{b}}$ | $\hat{b}$ | | $t_{\hat{b}}$ | $\hat{b}$ |
| **Panel A. Total GHG emissions** | | | | | | | | | | |
| | | -19.522* | -1.494 | | | | | | | |
| Club 1 | EE, PL, LV, BG, IE | 0.553 | 0.251 | 0.125 | | | | | | |
| Club 2 | CY, DE, PT | 0.344 | 0.073 | 0.036 | Club 1+2 | -0.181 | -0.060 | | | |
| Club 3 | NL, LT, AT | 1.168 | 1.026 | 0.513 | Club 2+3 | 0.559 | 0.128 | Club 1+2+3 | -0.564 | -0.163 |
| Club 4 | CZ, SI, SK, HU, FR, HR, LU, MT, SE, FI, BE, ES | 1.596 | 1.021 | 0.510 | Club 3+4 | -0.336 | -0.137 | Club 2+3+4 | -1.654* | -0.385 |
| Club 5 | RO, IT, DK, EL, UK | -1.384 | -1.231 | -0.615 | Club 4+5 | -3.610* | -0.662 | | | |
| **Panel B. GHG emissions in ETS sectors (stationary installations)** | | | | | | | | | | |
| | | -6.745* | -2.404 | | | | | | | |
| Club 1 | EE, NL | 1.711 | 2.392 | 1.196 | | | | | | |
| Club 2 | BG, CY, PL, DE, AT, SE, PT, CZ, SK, SI, IE, LV, FI, ES | 1.131 | 0.183 | 0.091 | | | | Club 1+2 | -2.827* | -0.502 |
| Club 3 | HU, FR, HR, EL, BE, DK | 0.126 | 0.100 | 0.050 | Club3 + IT | -0.291 | -0.208 | Club 2+3 | -3.756* | -0.641 |
| Club 4 | RO, LT | 1.034 | 1.986 | 0.993 | | | | Club 3+4 | -5.708* | -1.972 |
| Diverge | IT, LU, UK, MT | | | | | | | | | |
| **Panel C. GHG emissions in ESD sectors** | | | | | | | | | | |
| | | -17.727* | -2.718 | | | | | | | |
| Club 1 | PL, LV, RO | 0.553 | 0.251 | 0.125 | Club1 +LT | -1.562 | -0.844 | Club1+2 | -0.709 | -0.166 |
| Club 2 | CY, CZ, BG, DE, EE, HU, IE | 0.344 | 0.073 | 0.036 | | | | Club 2+3 | -4.303* | -1.118 |
| Club 3 | SI, HR, BE, FI, AT, FR, SK, PT, LU, ES | 1.168 | 1.026 | 0.513 | | | | Club 3+4 | -12.631* | -1.414 |
| Club 4 | NL, DK, IT, UK, SE | 1.596 | 1.021 | 0.510 | Club4 +EL | -2.951* | -0.732 | | | |
| Diverge | MT, LT, EL | | | | | | | | | |

* Rejection of the null of convergence at 5% significance.
Belgium (BE), Bulgaria (BG), Czech Republic (CZ), Denmark (DK), Germany (DE), Estonia (EE), Ireland (IE), Greece (EL), Spain (ES), France (FR), Croatia (HR), Italy (IT), Cyprus (CY), Latvia (LV), Lithuania (LT), Luxembourg (LU), Hungary (HU), Malta (MT), Netherlands (NL), Austria (AT), Poland (PL), Portugal (PT), Romania (RO), Slovenia (SI), Slovakia (SK), Finland (FI), Sweden (SE), United Kingdom (UK).

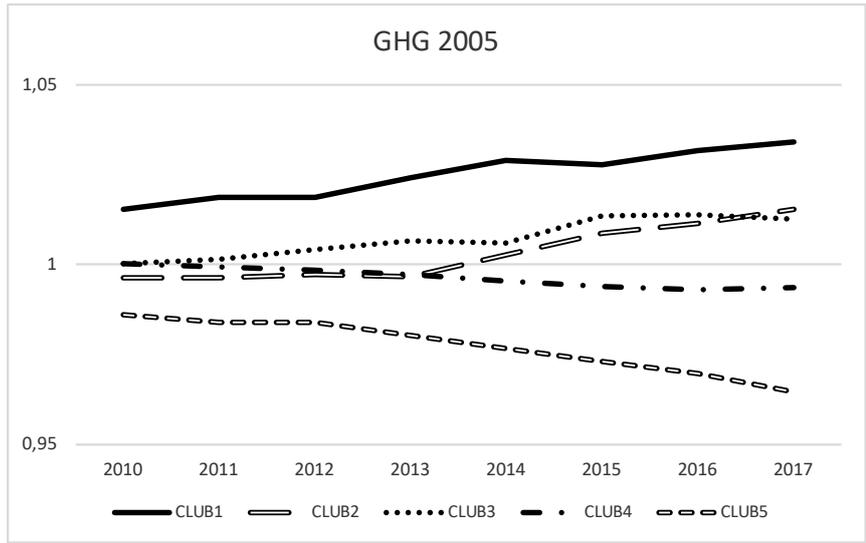

Panel A

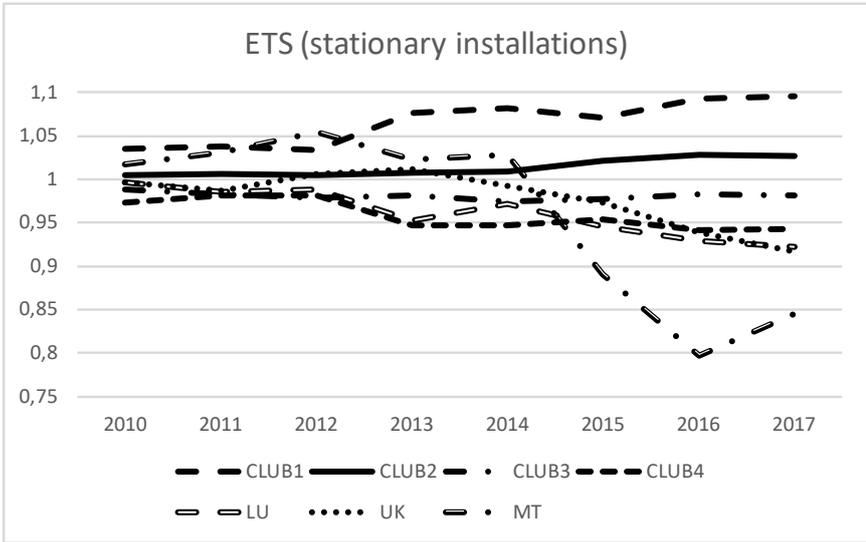

Panel B

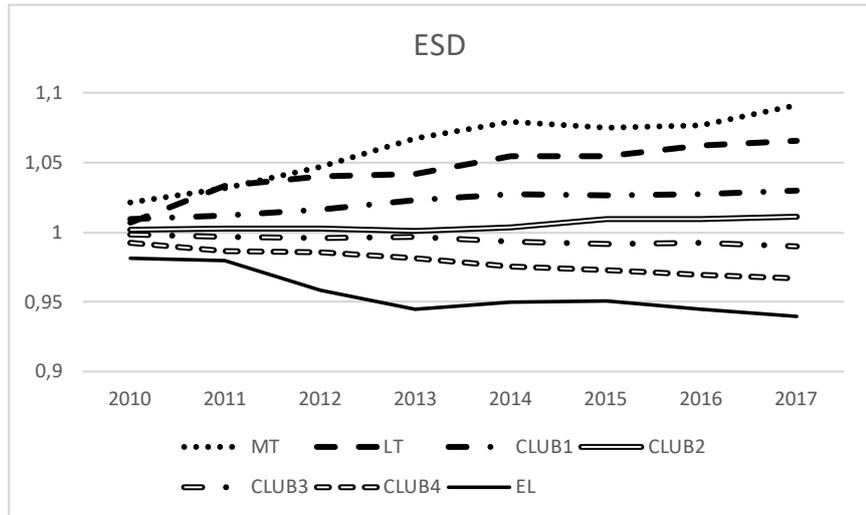

Panel C

**Figure 2. Relative transition paths across GHG emissions clubs (Global, ETS and ESD). 2005-2017**

**Table 3. Comparison among clubs in GHG emissions. 1990-2017 vs. 2005-2017**

|  |  | GHG 1990 | | |
|---|---|---|---|---|
|  |  | High (Club1) | Medium (Club2+3) | Low (Club4) |
| G H G 2 0 0 5 | High (Club1+2+3) | CY | EE, PL, BG, PT, IE, AT, NL | DE, LV, LT |
|  | Medium (Club4) | MT | FR, HR, SI, LU, FI, ES | BE, SE, HU, CZ SK |
|  | Low (Club5) |  | IT, EL | DK, UK, RO |

Belgium (BE), Bulgaria (BG), Czech Republic (CZ), Denmark (DK), Germany (DE), Estonia (EE), Ireland (IE), Greece (EL), Spain (ES), France (FR), Croatia (HR), Italy (IT), Cyprus (CY), Latvia (LV), Lithuania (LT), Luxembourg (LU), Hungary (HU), Malta (MT), Netherlands (NL), Austria (AT), Poland (PL), Portugal (PT), Romania (RO), Slovenia (SI), Slovakia (SK), Finland (FI), Sweden (SE), United Kingdom (UK).

In order to facilitate a more in-depth examination of the 2005-2017 period, a separate analysis of the evolution of emissions under ETS and ESD will be conducted in the following subsections.

**4.2.1. Clubs of convergence in ETS emissions**

The ETS is the world´s first and biggest carbon market. It aims at limiting the emissions from large point sources (mostly power and heat production and industrial installations) and other activities like cement, iron and steel production, and oil refining. Since 2012, the EU-ETS also encompasses emissions from aviation. Overall, it covers around 45% of the EU´s GHG emissions.

Total emissions have been in a downward trend in recent years. Thus, in year 2017, emissions from stationary installations under the EU-ETS were 26% below 2005 levels, mainly due to decreases in the emissions linked to power generation. This is quite a drop, as the EU-ETS target was to reduce emissions by 21% between years 2005 and 2020; moreover, forecasts submitted by the Member States in 2017 indicate that emissions from stationary installations are projected to decrease by 5% throughout the 2017-2020 period and by 7% between 2020 and 2030, with energy industries being the main driver of that reduction.

However, the evolution of the aviation sector reveals a very different pattern, with 22% growth in 2017 in relation to 2013 (Table A.2) and an increase in its share to around 4% of total EU28 GHG emissions and 13% of the emissions from transport.

Since data from aviation are only available from 2012 on, and with the aim of obtaining a homogeneous time series, emissions from that sector are not included in this study. In addition, in order to more accurately reflect the current scope of the EU-ETS (third trading period 2013-2020) and to ensure that the series analyzed were time consistent, we included the verified emissions from stationary installations plus estimates of emissions and allowances (EEA data) for the 2005–2012 period. (Qualitatively similar results, only including *verified* emissions, are available from the authors upon request).

The results for stationary installations (Table 2 and Figure 2, panel B) lead us to reject the null of convergence, with the following 4 convergence clubs being detected by the algorithm:

- Club 1 has a growing trend in its transition path and is integrated by Estonia and the Netherlands. The only country in the EU-28 that increased its ETS emissions in 2017 as compared with 2005 is Estonia, that heavily relies on its -highly emissions intensive- shale oil resources. For its part, Dutch ETS emissions in 2017 were similar to those of 2005, perhaps as a consequence of the fact that its local electricity system mostly uses fossil fuels for power generation, with the north of the country being rich in that resource.

- Club 2 is the largest group and had a transition path with values around the mean until year 2014, although showing a tendency to separate thereafter.

- Club 3 is characterized by a transition path below 1. The merging algorithm allows Italy to be included in this club, although convergence would be weak in that case.

- The most favorable evolution is observed for Romania and Lithuania (Club 4), in addition to Luxembourg, Malta and the United Kingdom (which diverge).

Table A.2 reports data on aviation emissions. As commented above, since year 2013 emissions increased by 22% in the EU-28, with the largest increases observed in Ireland. Certainly, aviation deserves specific attention when monitoring the evolution of GHG emissions.

**4.2.2. Clubs of convergence in ESD emissions**

The ESD sets national emissions targets (based on the relative wealth of each member state, measured by GDP per capita) for year 2020 expressed as percentage changes from 2005 levels. Targets include emissions from most sectors and activities –like road transport, buildings, agriculture, non-ETS industrial installations, and waste management-- not covered by the EU-ETS, which altogether amount to about 55% of EU's emissions.

Under the EU´s ESD of 2009, the Member States taken together were to achieve an overall emissions reduction of 10% by 2020 in relation to 2005. However, under the new Effort Sharing Regulation (adopted in 2018), EU-wide emissions are to be reduced by 30% by 2030 as compared with 2005.

Although 2017 was the third year in a row that ESD emissions increased, they still remained 10% below 2005 levels, so the ESD emissions reduction since that year was smaller than that experienced in the sectors under the EU-ETS. Certainly, that slower decrease was due to the transport component, despite the mitigating effect of the other components. The transport sector is the largest contributor (with a weight exceeding 33%) to emissions under the ESD, with increases observed since year 2015 following a drop between 2007 and 2014. The causes of that rise include increases in energy consumption (deriving from economic growth) in the road transportation sector, the expansion of tourism, rises in motorization rates, and the tendency to live in suburban areas.

Buildings are the second contributor, with a share around 25%. Emissions in that sector fell along with the decreases in consumption, with the older Member States of Northwest and Central Europe being the main drivers of those reductions.

Finally, in contrast to sectors in the EU-ETS (which are regulated at EU level), the policies and measures to limit emissions from the ESD sectors are responsibility of the member states, with a large concentration of emissions in a few countries. In this vein, in year 2017 six countries (Germany, the United Kingdom, France, Italy, Spain, and Poland) accumulated around 70% of total EU-ESD emissions.

The null of overall convergence in ESD emissions is also rejected (Table 2 and Figure 2. Panel C). This is barely striking, since the country targets for year 2020 in that chapter range between 20% drops and 20% increases.

Malta and Lithuania diverge as consequence of rises in their transport and building emissions. It is remarkable that these are among the states with targets allowing them to increase emissions in 2020 in relation to 2005, although both countries (especially Malta) are a far cry from their respective targets. The opposite pattern appears in Greece, that decreased ESD emissions about 30% as a consequence of large drops in all their components, including transport (-20%) and buildings (-59%).

- Clubs 1 and 2 exhibit an evolution of emissions above the mean of the EU and rising transition paths. Most countries in these clubs show positive targets, with three concerning exceptions: Cyprus, Germany and Ireland. The latter two states have ESD indexes slightly over the mean of the EU-28, with a rebound in recent years due to increases in transport emissions. It is not surprising that these three countries are among those having national projections pointing to emissions that exceed their Annual Emission Allocations (AEA) by 2020 (EEA, 2018).

A key factor explaining the inclusion of Poland in Club 1 is the strong rise in emissions in its transport (+68% in the period, the highest in the EU-28) and building sectors. Significant expansions were also observed in the emissions from the transport sector in other members of Clubs 1 and 2, as Romania, Bulgaria, and Estonia (the last two ones also experienced substantial increases in agriculture).

- Clubs 3 and 4 show transition paths below the EU mean and are mainly integrated by states with negative (i.e., decrease) goals, but also by a small group of countries (Slovenia, Croatia, Slovakia, and Portugal) whose targets are positive.

Club 3 (transition path near 1) includes the rest of the countries that might exceed their AEAs by 2020: Austria, Belgium, Finland, and Luxembourg (Table A.2). For its part, Club 4 has the most favorable evolution in ESD emissions, with transition paths that --in accordance with their ambitious goals-- tend to separate from the EU mean. All sectors under the ESD contributed to that decrease (the only exception is Netherlands, in agriculture). It is also remarkable the inclusion of Italy in the club, since that country reduced emissions around 19% in the period, by far exceeding its objective. In addition, two club members (Sweden and Italy) are among the exceptions to the increasing trend in transport sector emissions since 2014. In the case of Sweden this would fit with its high share (38.6%, the highest in the EU-28 and multiplying by 6 since 2005) of renewable energy in the gross final energy consumption of the transport sector. Morales-Lage *et al.* (2019) also stress the divergent behavior of the Swedish transportation sector, in their convergence analysis for the energy subsectors.

### 4.2.3. Joint analysis

Finally, in order to draw some conclusions on those sectors where additional efforts are required, Table 4 below summarizes the results for global, ETS and ESD emissions.

**Table 4. Clubs for GHG, ETS (stationary installations) and ESD emissions**

| | | ETS (stationary installations) | | | | ESD | | |
|---|---|---|---|---|---|---|---|---|
| | | High (Club1) | Medium-High (Club2) | Medium-Low (Club3+IT) | Low (Club4+ LU+UK+ MT) | High (MT+LT+ Club1+2) | Medium (Club3) | Low (Club4+EL) |
| GHG 2005 | High (Club1+2+3) | EE, NL | BG, PL, DE, PT, CY, AT, IE, LV | | LT | EE, PL, LV, BG, IE, CY, DE, LT | PT, AT | NL |
| | Medium (Club4) | | SE, SK, CZ, SI, FI, ES | BE, HR, HU, FR | LU, MT | CZ, MT, HU | SI, SK, FR, HR, LU, FI, BE, ES | SE |
| | Low (Club5) | | | IT, DK, EL | UK, RO | RO | | IT, DK, EL, UK |

Belgium (BE), Bulgaria (BG), Czech Republic (CZ), Denmark (DK), Germany (DE), Estonia (EE), Ireland (IE), Greece (EL), Spain (ES), France (FR), Croatia (HR), Italy (IT), Cyprus (CY), Latvia (LV), Lithuania (LT), Luxembourg (LU), Hungary (HU), Malta (MT), Netherlands (NL), Austria (AT), Poland (PL), Portugal (PT), Romania (RO), Slovenia (SI), Slovakia (SK), Finland (FI), Sweden (SE), United Kingdom (UK).

Among the countries (labelled as "high" in Table 4) whose evolution of emissions is above the mean of the EU, most of them (Germany, Ireland, Estonia, Poland, Latvia, Bulgaria, and Cyprus) were also above average in both ESD and ETS emissions. However, the main driver of the inclusion of the Netherlands in Club 1 might be ETS emissions, while the opposite behavior is observed in Lithuania.

At the other end, among the states with emissions evolving below the mean, the United Kingdom had a favorable evolution in both ETS and ESD emissions, in contrast to Romania (probably as a consequence of its significant increases in the transport sector). The good progress of the ESD sectors may have been a relevant factor in the inclusion of countries like Italy, Greece and Denmark in Club 5.

Finally, Sweden had a much more favorable evolution in ESD than in stationary installations emissions. Regarding this, it must be remembered that emissions from aviation (which have a non-negligible weight in Swedish emissions) are not included in our analysis. It is also remarkable the inclusion in the same club of countries (e.g., Slovakia, Finland and Spain) that evolved in similar ways, while having fairly heterogeneous characteristics. Belgium and France are also in the same clubs for both ETS and ESD sectors.

5. **Conclusions**

Climate change is a reality and international organizations and goverments have been for years raising awareness of it and trying to reverse the process. The UE is not an exception in this regard and has developed policies to fight climate change, monitoring their deployment and efficiency through a range of indicators.

This paper has focused on the evolution of an indicator -GHG emissions- connected with climate mitigation. We have analyzed its convergence among the EU-28 countries. This kind of analysis is most relevant when evaluating the efficiency of environmental policies, since implementing common enviromental policies should be easier when countries converge. Within the framework of strategies like Europe 2020 it is important to assess if all the EU-28 countries converge to the same steady state, and in the event of failure, to identify groups of countries converging to similar steady states in their growth paths.

The methodology we have relied on was convergence club analysis. The hypothesis of convergence in the GHG emissions index is rejected for the 1990-2017 period, with four clubs initially identified by the algorithm. Among the countries that reduced their emissions above the UE-28 average were the United Kingdom, Germany, Belgium, Sweden, Denmark, and some ex-comunist states like Hungary, the Czech Republic, Slovakia, Romania, Latvia and Lithuania. A decoupling between growth and emissions is observed in that group, with some countries experiencing strong growth troughout the study period. All those nations underwent restructuring of their energetic systems, from high-carbon fossil fuels to low-carbon energy technologies, and increased their use of renewables and energy efficiency.

Some countries in the study were not Member States in 1990, which led us to a separate analysis for the 2005-2017 subperiod. A somewhat different scenario emerges in that case: among the new members, only Romania remains in the group having the most favorable evolution; among the older members, only the United Kingdom and Denmark. Two countries now join the club, namely Italy and Greece, both affected by the crisis.

Individual analysis of ETS (stationary installations) and ESD emissions allowed us to detect countries where the efforts to control emissions in some sectors should be redoubled. The countries placed in clubs of convergence above the mean of the EU in both ETS and ESD are Estonia, Bulgaria, Poland, Latvia, Cyprus, Germany, and Ireland. On the other hand, Lithuania and Romania show a favorable evolution in ETS, with gross inland consumption reduced by a similar percentage, while (especially in the case of Lithuania) their strong growth resulted in above-average evolution in the ESD sectors. An opposite pattern is observed in the Netherlands and, to a lesser extent, Sweden. Finally, Italy, Greece, Denmark, and mainly the United Kingdom are all in below-average convergence groups both in ESD and ETS sectors.

The above results have some policy implications. First, our finding of multiple clubs, reinforced by persistence of the results for the various periods and sectors analyzed, may signal the convenience of devising differentiated, more specific policies and regulations oriented to reduce heterogeneity. Those countries located in worse performing clubs would need special attention, with both more effective public initiatives to incentivize fulfillment of their goals and an accelerated implementation of new policies: different countries would demand different measures and targets to be adopted.

ETS emissions from Member States' stationary installations decreased around 25% in 2017 as compared with 2005. That drop, which occurred in spite of the increase in emissions from aviation, may be explained by several factors including the effectivity of the European carbon market and reductions of production as a consequence of the economic crisis (that severely hit all energy-intensive industries, with the exception of the chemical sector), but also by drivers like the development of low-carbon, breakthrough technologies for energy-intensive industry processes, enabled by various European R&D programmes and private initiatives. Therefore, profiling energy intensive industries by favoring the development and implementation of new and better technological options, promoting a circular economy and looking for new business models, in a framework that is well-regulated, transparent, and endowed with modern infrastructures, should all lead to a remarkable reduction in emissions.

Large mitigation potentials also exist in the multiplicity of sectors covered by the ESD, especially in road transport. Although the battery of ESD national emission targets is certainly useful, it might be strenghtened through further measures. In this regard, initiatives oriented to long-term decarbonisation would include improving transport management, reducing the demand for transport, and encouraging a shift towards more environment-friendly modes, in addition to enhancing vehicle energy efficiency, electric and hybrid vehicles, and the use of liquid biofuels both for the current stock of vehicles with internal combustion engines and for those transport modes where electrification is unfeasible at the present time. In addition to increasing the share of alternative fuels, integrated measures addressing both production and consumption may be necessary in the long run in order to curb the expected increase in transport demand and reduce its emissions. With transport being one of the key sectors to meet the EU's commitments under the Paris Agreement, such initiatives as the European Strategy for Low-Emission Mobility, the Seventh Environment Action Programme, and the EU plans on Accelerating Clean Energy Innovation make up important milestones on the way to reducing GHG emissions in that subsector. Other potentially useful measures would include making the organization of the transport system more efficient (with smart solutions for pricing, telecommuting, and logistic), improving and promoting the use of public transport, a modal shift from cars to human-powered vehicles like the bicycle, and encouraging the electrification of transport (favoring the use of vehicles using electricity, as well as hydrogen, renewable natural gas, and advanced biofuels). In the specific case of biofuels, market uncertainties arising from changes (oriented to strengthen their sustainability criteria) in EU's policy after 2018 have been relatively unfavorable, and governments should learn from past experiences, providing a safe and stable legal framework to all stakeholders.

Efforts to reduce emissions in the building sector have also brought about some results, although further improvements are needed in the direction of using renewable energies for heating/cooling, support for retrofitting buildings, and active investment and research in efficient systems. In the case of farms, those efforts should also include the conversion of livestock manure to biogas.

Renewable energy sources play a vital role in order to achieve the five dimensions of the Energy Union, namely security of energy supply, market integration, energy efficiency, decarbonization, and innovation. Over the last decades the EU has been at the forefront of global renewable energy deployment, through adoption of long-term targets and support to policy measures, with the EU-28 having duplicated its share of renewable energy in the 2004-2017 period. Nevertheless, over the coming years, the EU should increase its efforts in fields like solar photovoltaic and wind power generation, progress in deployment of renewables for heating and cooling, and use of renewable transport options. On the road to decarbonizing the system by expanding the use of renewable sources, technology research and investment will surely be key means to ensure the objectives of the Energy Union. Moreover, those investments may also derive substantial benefits in terms of growth, employment, industrial competitiveness, balance of trade, and creation of a new industrial base around the renewables sector. For instance, much higher levels of final electricity demand are expected if industrial low-carbon technologies are deployed across the EU (especially in some Eastern members). Under proper conditions, a transition to higher levels of electrification has the potential to create a virtuous cycle between renewable energy and industrial transition. The European Commission is actively promoting new ways to transfer the results from those research projects into the market. In this regard, the dialogue and interaction -between now and year 2030- among the European Commission and the Member States around their National Energy and Climate Plans will surely be crucial.

Finally, the current economic crisis, deriving from the covid-19 pandemic and with significant consequences on energy consumption, may well be an opportunity for a rapid transition towards a green economy. Initiatives as the Recovery and Resilience Facility (EU Commission; May 2020), which will make €672.5 billion in loans and grants available to support reforms and investments undertaken by Member States, will be a great opportunity both to mitigate the negative effects of the pandemic and to make EU economies more resilient, sustainable, and in a better position for the green transition.

Apergis, N., Payne, J.E., 2017. Per capita carbon dioxide emissions across U.S. states by sector and fossil fuel source: Evidence from club convergence tests. Energy Economics. 63, 365-372.

Apergis, N., Payne, J.E., Topcu, M., 2017. Some empirics on the convergence of carbon dioxide emissions intensity across US states. Energy Sources Part B: Economics, Planning and Policy. 12(9), 831-837.

Azariadis, C., Drazen, A., 1990. Threshold externalities in economic development. Quarterly Journal of Economic Development. 105, 501-526.

Balsalobre-Lorente, D., Shahbaz, M., 2016. Energy consumption and trade openness in the correction of GHG levels in Spain. Bulletin of Energy Economics. 4, 310-322.

Barro, R.J., Sala-i-Martin, X. 1990. Economic growth and convergence across the United States. NBER Working Paper No. W3419.

Baumol, W.J. 1986. Productivity growth, convergence, and welfare: what the long-run data show. American Economic Review. 76, 1072–1085.

Bilgili, F., Ulucak, R., 2018. Is there deterministic, stochastic and/or club convergence in ecological footprint indicator among G20 countries? Environmental Science and Pollution Research. 25 (35), 35404-35419.

Burnett, J.W., 2016. Club convergence and clustering of U.S. energy related $CO_2$ emissions. Resource and Energy Economics. 46, 62-84.

Camarero, M., Picazo-Tadeo, A.J., Tamarit, C., 2013. Are the determinants of $CO_2$ emissions converging among OECD countries? Economics Letters. 118, 159-162.

Carlino, G.A., Mills, L.O., 1993. Are US regional incomes converging?: A time series analysis. Journal of Monetary Economics. 32, 335–346.

Emir, F., Balcilar, M., Shahbaz, M., 2019. Inequality in carbon intensity in EU-28: analysis based on club convergence. Environmental Science and Pollution Research. 26(4), 3308-3319.